\documentclass[%
 aip,
 amsmath,amssymb,
preprint,%
]{revtex4-1}

\usepackage{graphicx}
\usepackage{dcolumn}
\usepackage{bm}

\usepackage[utf8]{inputenc}
\usepackage[T1]{fontenc}
\usepackage{mathptmx}
\usepackage{etoolbox}

\usepackage{xcolor}

\makeatletter
\def\@email#1#2{%
 \endgroup
 \patchcmd{\titleblock@produce}
  {\frontmatter@RRAPformat}
  {\frontmatter@RRAPformat{\produce@RRAP{*#1\href{mailto:#2}{#2}}}\frontmatter@RRAPformat}
  {}{}
}%
\makeatother
\begin{document}

\preprint{AIP/123-QED}

\title[SW in FiMs near the AMC Temperature]{Spin Waves in Ferrimagnets near the Angular Magnetization Compensation Temperature: A Micromagnetic Study}
\author{Luis Sánchez-Tejerina}
\affiliation{ 
Department of Electricity and Electronics, University of Valladolid, Spain.
}
\author{David Osuna Ruiz}
\affiliation{ 
Departament of Electrical, Electronical and Communications Engineering, Public University of Navarra, Spain.
}
\author{Eduardo Martínez}
\affiliation{ 
Department of Applied Physics, University of Salamanca, Spain.
}
\author{Luis López Díaz}
\affiliation{ 
Department of Applied Physics, University of Salamanca, Spain.
}
\author{Óscar Alejos}
\email{oscar.alejos@uva.es.}
\affiliation{ 
Department of Electricity and Electronics, University of Valladolid, Spain.
}

\date{\today}

\begin{abstract}
Spin wave propagation along a ferrimagnetic strip with out-of-plane magnetization is studied by means of micromagnetic simulations. The ferrimagnetic material is considered as formed by two antiferromagnetically coupled sub-lattices. Two critical temperatures can be defined for such systems: that of magnetization compensation and that of angular momentum compensation, both different due to distinct Landé factors for each sub-lattice. Spin waves in the strip are excited by a spin current injected at one of its edges. The obtained dispersion diagrams show exchange-dominated forward volume spin waves. For a given excitation frequency, Néel vector describes highly eccentric orbits, the eccentricity depending on temperature, whose semi-major axis are oriented differently at distinct locations on the FiM strip.
\end{abstract}

\maketitle

Magnon propagation in antiferromagnetic (AFM) and/or ferrimagnetic (FiM) has drawn attention recently and is the subject of very recent works\cite{Chen:23,Rezende:19,Cheng:16,Puliafito:19,Xu:19,Millo:23,Yanes:20,Wang:15,Safin:20} due to the variety of benefits as opposed to using ferromagnetic (FM) materials, such as greater speeds, marginal sensitivity to external fields and the wider range of materials available, many being electrical insulators. Spin wave propagation in insulating AFM materials connects magnonics with spintronics since spin current (a key concept in spintronics) is mainly transported by magnons in such materials, due to the absence of conducting electrons as opposed to FM materials, where electrons are the main spin carriers. The use of insulators also points to the importance of thermal effects due to local heating. The dynamics of spins in AFM materials have been widely investigated theoretically. For example, characterising the fundamental $k = 0$ resonance modes \cite{Rezende:19, Cheng:16, Puliafito:19}, propagating magnons\cite{Xu:19,Millo:23} or AFM domain walls (DWs) and their interactions with magnons\cite{Yanes:20, Wang:15}, but always focused on their fundamental physics, as AFM magnonics is a field still in its infancy \cite{Rezende:19, Safin:20, Barman:21}. Additionally, FiM materials at the angular momentum compensation point ($T_A$) can mimic the dynamical response of AFM\cite{Caretta:18, Siddiqui:18, Cutugno:21} while being easier to manipulate and detect by electric and/or optical means\cite{Ueda:2016, Fleischer:18}.

In this work, micromagnetic simulations have been used to investigate the propagating high-frequency ($\approx \text{THz}$) AFM modes in a FiM strip at different temperatures. Simulations consider FiMs as formed by two sub-lattices (indexes $i=1,2$ will be used to refer to each sub-lattice). Accordingly, a pair of coupled Landau-Lifshitz-Gilbert equations are used to model the material:\cite{Martinez:20b}
\begin{equation}
\dot{\mathbf{m}}_i=-\gamma_i\mathbf{m}_i\times\mathbf{H}_{\text{eff},i}+\alpha_i\mathbf{m}_i\times\dot{\mathbf{m}}_i+\tau_{\text{SOT},i}\qquad \left(i=1,2\right)\text{,}   
\end{equation}
where $\mathbf{m}_i$, $\gamma_i$ and $\alpha_i$ are the local orientation of the magnetization, the gyromagnetic factor, and the damping parameter for each sub-lattice, respectively. $\mathbf{H}_{\text{eff},i}$ is the effective field including all relevant interactions within the system, and $\tau_{\text{SOT},i}$ corresponds to the spin-orbit torque (SOT) due to spin currents. In our model, the FiM is formed by computational elementary cells, in a finite difference scheme. We have two magnetic moments within each cell, one for each sub-lattice, and both sub-lattices are coupled by an interlattice exchange interaction (see details in Ref.~\cite{Martinez:20b}).

\begin{figure*}
\includegraphics[scale=1]{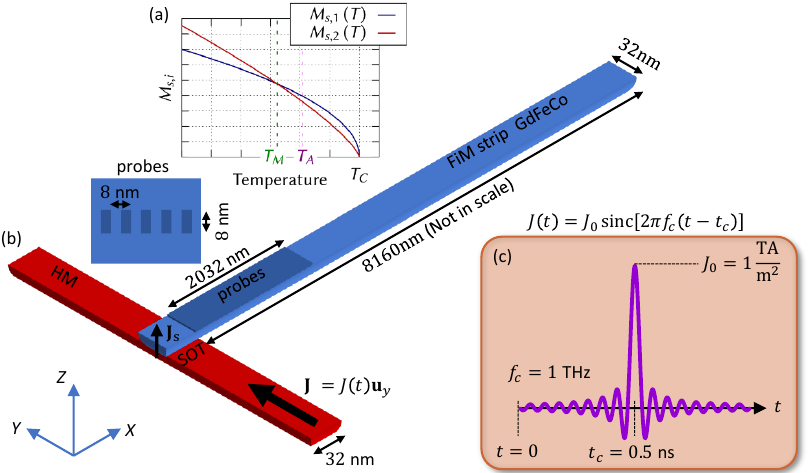}
\caption{\label{fig:01}Geometry and properties of the modeled device: (a) Temperature dependence of the saturation magnetization for each sub-lattice, (b) dimensions of the FiM strip and of the exciting spin current injection area, (c) time dependence of the exciting current to obtain the dispersion diagrams.}
\end{figure*}

Simulations have been carried out by means of a homemade code\cite{Alejos:18} implemented on graphic processing units. Details of the modeled device are depicted in Fig.~\ref{fig:01}. FiM strip parameters considered are those that can be found in the literature for a prototypical FiM as GdFeCo\cite{Caretta:18}. Saturation magnetization for each sub-lattice varies with temperature according to the law $M_{s,i}=M_{s,i}^0\left(1-\frac{T}{T_C}\right)^{b_i}$, $T_C=450~\text{K}$ being the Curie temperature, $M_{s,i}^0$ being the saturation magnetization at $0\text{K}$: $M_{s,1}^0=1.71 \frac{\text{MA}}{\text{m}}$ $M_{s,2}^0=1.4 \frac{\text{MA}}{\text{m}}$, and $b_1=0.76$, and $b_2=0.5$, resulting in the temperature dependence plotted in Fig.~\ref{fig:01}(a). According to these parameters, magnetization compensation occurs at $T_M=241~\text{K}$. The rest of significant parameters are: intralattice exchange $A_i=70\frac{\text{pJ}}{\text{m}}$, interlattice exchange $B_{ij}=-9\frac{\text{MJ}}{\text{m}^3}$, damping constant $\alpha_i=0.001$, anisotropy constant $K_{u,i}=1.4 \frac{\text{MJ}}{\text{m}^3}$, Dzyaloshinskii-Moriya constant $D_i=0.12\frac{\text{mJ}}{\text{m}^2}$, and spin-Hall angle $\theta_{SH}=0.15$. Finally, each sub-lattice is characterized by distinct Landé factors: $g_1=2$, $g_2=2.05$, which result in a temperature of angular momentum compensation $T_A=260\text{K}$, slightly greater than $T_M$.

The device consists of a FiM strip and a current line of a heavy metal (HM) running perpendicular to the strip at one end as shown in Fig.~\ref{fig:01}(b). The FiM material is initially uniformly magnetized in the out-of-plane direction. The FiM strip area is $8192~\text{nm}\times 32~\text{nm}$, and it is 6-nm thick, while the HM strip is 32-nm wide. This defines a $32~\text{nm}\times 32~\text{nm}$ SW excitation area on the leftmost end of the FiM strip due to an electric current flowing through the HM line, then exciting spin waves via SOT. Beside this area, a set of 254 probes are regularly spaced on the FiM strip to monitor the value of the local magnetization, i.e., Néel's vector $\mathbf{n}=\mathbf{m}_1-\mathbf{m}_2$. Probes are $3~\text{nm}\times 8~\text{nm}$ in size and 5-nm distant from each other, then occupying a total length of $2032~\text{nm}$. 

A spin current arising from an electric current carried by the HM strip is injected through the excitation area. This spin current is polarized along the $\pm\mathbf{u}_x$ direction depending on the instantaneous direction of the electric current along the HM strip. Figure~\ref{fig:01}(c) presents the time dependent current considered to achieve a uniform excitation across a desired frequency range in the dispersion diagram. In this first case, an (unnormalized) sinc-shaped electric current pulse has been used, $J\left(t\right)=J_0\mathrm{sinc}\left[2\pi f_c\left(t-t_c\right)\right]$, where $f_c$ is the excitation cut-off frequency, which was set to $1~\text{THz}$, $J_0$ is the pulse amplitude, initially set to $1~\frac{\text{TA}}{\text{m}^2}$, and $t_c$ is a delay time which has been chosen as one half of the total simulation time, set to $1~\text{ns}$. Using this activation, we ensure that each mode is equivalently fed. 
This current is sufficiently small to remain in the linear regime of activation of damped, stable oscillations\cite{Cheng:16} and to avoid any static changes in the AFM phases
of the sample. The delay time also provides a reasonable offset to the peak of the pulse, allowing a gradual increase of the amplitude from the beginning of the simulation. However, when later analyzing the time evolution of the magnetic signal, a monochromatic wave (MW) excitation is to be applied with a current at a specific frequency $f_0$, i.e., $J\left(t\right)=J_0\sin\left(2\pi ft\right)$. Equivalently to the previous case, $J_0$ is chosen to be $1~\frac{\text{TA}}{\text{m}^2}$ to remain in the same modes regime and to obtain a good magnetization contrast for $\mathbf{m}_{1,2}\left(t\right)$. Finally, temperature is taken here as an additional parameter, i.e., a way to balance the value of the saturation magnetization of both sub-lattices to achieve both magnetization compensation and angular momentum compensation. This means that only coherent magnons are considered in our simulations.


SW spectra in AFM materials have revealed ultra-high frequency (in the THz regime), field-dependent modes that are related to each of the sub-lattices (strongly coupled by exchange) usually called ‘optic’ or higher frequency mode, and ‘acoustic’ or lower frequency mode
\cite{Rezende:19,Cheng:16}. For each mode, there are opposite senses of precession for each sub-lattice, either right-handed (RH) or left-handed (LH). The AFM resonance frequencies ($k = 0$ magnons) depend on the different equilibrium states of the coupled sub-lattices, or so-called AFM phases\cite{Rezende:19}, which can also be tuned by external fields. These modes can be locally excited via spin currents $\mathbf{J}_s$, as represented in Fig.~\ref{fig:01}(b). The localized perturbation allows AFM magnons with $k \neq 0$ to propagate in the material via exchange interactions, consequently possessing very short wavelength. To investigate the SW propagation in the frequency domain, we analyze the Néel vector since it transports coherently excited magnons, such as those generated by $\mathbf{J}_s$\cite{Lebrun:18}.

\begin{figure*}[h]
\includegraphics[scale=1]{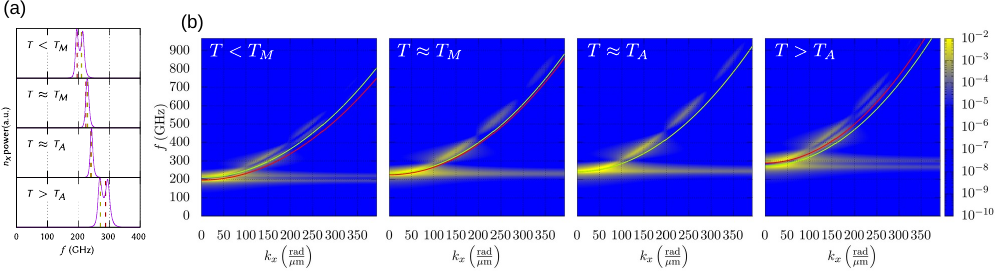}
\caption{\label{fig:02}(a) SW spectra (FFT intensity of $n_x$) and fundamental resonances at $k = 0$ for different strip temperatures, and (b) dispersion relations obtained for $n_y$ at four different temperatures in the strip ranging below $T_M$ to above $T_A$. The regularly spaced interruptions in the dispersion curves are related to the geometry of the excitation zone (see text).}
\end{figure*}

We focus on the in-plane components of the Néel vector $n_x$ and $n_y$ which account for the dynamic (differential) x-component and y-component respectively from the two sub-lattices at the closest position to the excitation region. Figure~\ref{fig:02}(a) shows the SW spectra (FFT intensity of $n_x$) and fundamental resonances at $k = 0$ for different strip temperatures ranging from below $T_M$ to above $T_A$, as a response to the sinc-shaped pulse current. The simulated (normalised) spectra are in good agreement with recent work characterising such modes in FiMs\cite{Haltz:22}, where the two distinctive peaks overlap at $T_A$. Similar results are obtained for $n_y$ (not shown), but with greater amplitude than $n_x$ at $T_A$, and generally wider peaks, suggesting an elliptical precession of $\mathbf{m}_{1,2}$ \cite{Rezende:19}. In contrast to $n_y$, $n_x$ is more attenuated at $T_A$ than at any other temperature. In particular, Fig.~\ref{fig:02}(b) shows the dispersion relations obtained for $n_y$ at four different temperatures in the strip ranging below $T_M$ to above $T_A$. The color scale is the same for all color plots, so intensities are directly comparable between different temperatures. The exchange-dominated character of the SWs is clear, in the form of a exchange-dominated $k^2$-dependent mode frequency\cite{Kalinikos:86}. Solid curves show the fitting to the data from Kalinikos and Slavin’s theory\cite{Kalinikos:86}, when each sub-lattice magnetization is modeled as of separate effective FM, i.e. as in a synthetic antiferromagnet (SAF), and $M_{S,i}(T_k)$. A Forward Volume SW (FVSW) configuration is assumed, which corresponds well to the initial out-of-plane configuration, where $\mathbf{m}_{1,2}$ are perpendicular to both $\mathbf{k}$ and the strip plane at equilibrium. When the AFM resonance frequencies ($k = 0$) for each sub-lattice (see Fig.~\ref{fig:02}(a)) are included as the \emph{cut-off} frequencies into the FVSW equations, the agreement with the simulated dispersion relations is remarkable (green curve for $i = 1$ and red curve for $i = 2$). 
Such dispersion relations can be analytically described as\cite{Rezende:19}
\begin{equation}
\omega_i=\gamma_i\left(H_{c,i}+\frac{2A_i}{\mu_0M_{S,i}}k_x^2\right)\qquad\left(i=1,2\right)\text{,}
\end{equation}
with $H_{c,i}=\sqrt{2H_{anis,i}H_{exch,i}+H_{anis,i}^2}$, $H_{exch,i}=\frac{\left|B_{ij}\right|}{\mu_0M_{S,i}}$ being the interlattice exchange field, and $H_{anis,i}=\frac{2k_{u,i}}{\mu_0M_{S,i}}$ being the anisotropy field. According to these expressions, the pair of dispersion curves merges as the temperature approaches $T_A$, as Fig.~\ref{fig:02}(b) confirms. It should be mentioned at this point that the dispersion curves are regularly interrupted for the various multiples of a certain $k_x\approx 100~\frac{\text{rad}}{\mu\text{m}}$. This modulation is caused by the spatially limited excitation zone $32~\mathrm{nm}$-wide. Note that at the edges of the excitation zone, the spin accumulation vanishes giving a half-wavelength of $\frac{\lambda}{2}=32~\mathrm{nm}$ corresponding to a $k$-vector with $k_x=\frac{2\pi}{\lambda}$.

\begin{figure}
\includegraphics[scale=1]{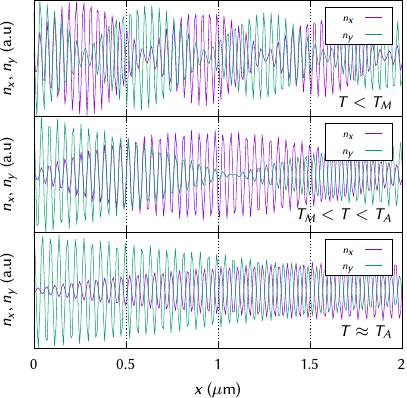}
\caption{\label{fig:03} Typical instantaneous values of $n_x$ and $n_y$ at the location of each probe. The graphs shows the results obtained at $t=1~\text{ns}$ at different temperatures: $T<T_M$, $T_M<T<T_A$ and $T\approx T_A$.}
\end{figure}

With the purpose of analyzing the time evolution of the magnetic signal under a MW excitation, the following results presented correspond to an excitation via a MW signal of frequency $f=360~\text{GHz}$. Although other frequency values have been considered, this one has been found to be optimal for revealing magnon propagation characteristics along the FiM strip. In particular, this value is slightly higher than the \emph{cut-off} frequency ($k=0$ magnons) for all of the temperatures considered. The intersection of this excitation frequency with the dispersion curves shows how the wave vector is in general different for each of the two sub-lattices. Accordingly, SWs travel with different phase velocities through each sub-lattice. This fact is illustrated in the graphs in Fig.~\ref{fig:03}. The plots show projections along the x- and y-directions of the N\'{e}el vector orbits. Plots are obtained at the end of the excitation time to ensure that the stationary regime is achieved. The coupled oscillations between both sublattices but with different $k_x$ generate patterns of nodes and anti-nodes of $n_x$ and $n_y$ along the FiM strip. All patterns possess anti-nodes of $n_y$ at the closest zone to the excitation area in agreement with the injected spin current. In addition to the oscillatory behavior of the two components, their amplitudes are also 
modulated by the progressive vanishing of the amplitude of the oscillations as the SWs travel along the strip. To get a clearer idea of the magnon propagation, a set of videos showing the evolution in time of the plots in Fig~\ref{fig:03} has been added as part of the supplementary material.

\begin{figure*}
\includegraphics[scale=1]{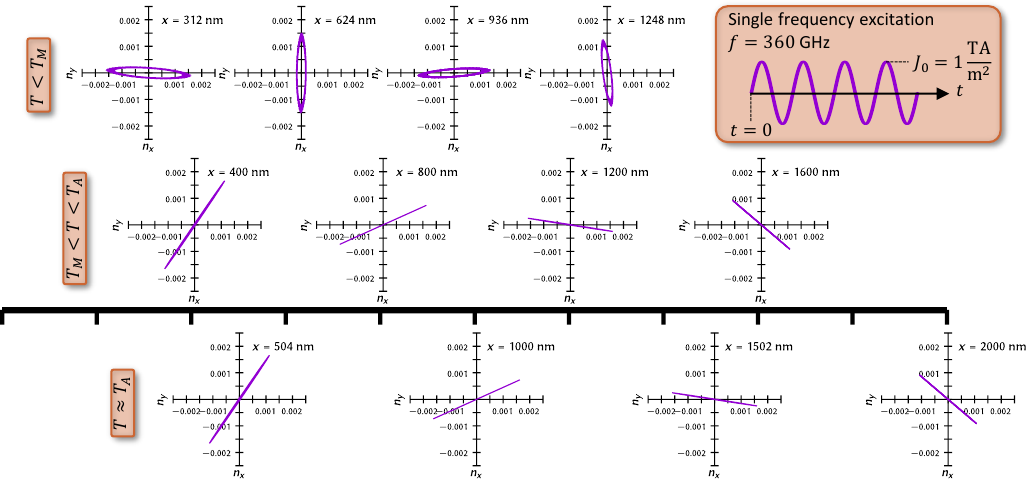}
\caption{\label{fig:04}Trajectories of the in-plane component of the Néel vector in the stationary regime under a MW excitation. Trajectories are obtained at different locations along the FiM strip and at different temperatures.}
\end{figure*}

We can further analyze the consequences of the phase velocity difference between the two sub-lattices. Fig.\ref{fig:04} represents the in-plane components of the N\'{e}el vector at different positions along the strip for three different temperatures: below $T_M$, in-between $T_M$ and $T_A$, and at approximately $T_A$. These trajectories are always elliptic, their eccentricity being determined by temperature and reaching a maximum at $T_A$. Fig.\ref{fig:04} also shows that the strip position plays no role in the eccentricity, though the semi-major axis takes different orientations depending on location. The N\'{e}el vector trajectory is composed of the two magnetization sub-lattice precessions. Because the phase velocity of each sub-lattice differs, the relative phase between the two precessions is modified from point to point, thus modifying the semi-major axis orientation. Consequently, different positions are subjected to different torques, not only by the amplitude fading but because of this reorientation effect. Nevertheless, the reorientation vanishes as the temperature approaches $T_A$, which can be a touchstone to detect this temperature for FiM materials under test. 

In conclusion, we have numerically explored the excitation and propagation of SWs along FiM strips as a function of temperature, below and above the angular and the magnetization compensation points. The analysis of results could be extrapolated to FiM materials with different compositions at room temperature. The fact that the eccentricity of the orbits can be controlled by temperature, in addition to the gradual rotation of the semi-major axis along the strip, provides an unprecedented local control of magnetization oscillations, which can be exploited to implement novel functionalities such as sensing in the emerging field of THz magnonics \cite{Barman:21}. Our results may also be relevant to spintronics for making very localized excitation positions given the sensitivity to distance of the different exerted torques along the magnetic strip. Besides, a potential reconfigurability can be envisaged by using controlled heat sources along the strip, such as laser pulses.



This work was supported by Projects No. SA114P20 from Junta de Castilla y León, No. PID2020117024GB-C41 funded by MCIN/AEI/10.13039/501100011033, and Project MagnEFi, Grant Agreement No. 860060, (H2020-MSCA-ITN-2019) funded by the European Commission.

The data that support the findings of this study are available from the corresponding author upon reasonable request.

\bibliography{mmag}

\end{document}